\documentclass[runningheads,a4paper]{llncs}

\usepackage{amssymb}
\usepackage{graphicx}
 
\usepackage{url}

\newcommand{\keywords}[1]{\par\addvspace\baselineskip
\noindent\keywordname\enspace\ignorespaces#1}

\begin{document}


\title{Domain Specific Distributed Search  Engine Based on Semantic P2P Networks }

\titlerunning{Distributed Search  Engine Based on Semantic P2P Networks}

%
%
\author{Lican Huang
 \\}

\authorrunning{Lican Huang}

\institute{Zhejiang Sci-Tech University , Domain Search Networking Technology Co., Ltd,\\
HangZhou, P.R.China, 310018  \\  LicanHuang@zstu.edu.cn; huang\_lican@yahoo.co.uk\\ }

%
%

\toctitle{Lecture Notes in Computer Science}
\tocauthor{Authors' Instructions}
\maketitle

\begin{abstract}
 This paper presents a distributed search engine based on semantic P2P Networks.
 The user's computers  join the domains in which  user wants to share information in  semantic P2P networks  which is domain specific virtual tree (VIRGO ). Each user computer contains search engine which indexes the domain specific information on local computer or Internet.  We can get all search information through  P2P message provided by all joined computers.  By  companies'  effort, we have implemented a prototype of distributed search engine, which demonstrates  easily retrieving  domain-related information  provided by joined computers .

\end{abstract}

\section{Introduction}

Today, the information such as documents, database ,etc  on the Servers or local machines are increasing rapidly.  How to retrieve the desired information is very important issue for the general users.  Although Google or Baidu supply index search engine nowadays , there still some big challenges. Firstly, the search results are not precise by Google or Baidu.  Although  PageRank\cite{PageRank} introduced by Google get more precise about keywords,
the search  engines  still response  much vague results, which are not desired ones for users.  Secondly, when the volume of information increases tremendously , the index search engine needs more computers, which will reach  the ceiling of computer power and economic feasibility.
Thirdly, there are billions of local computers which are behind NAT\cite{NAT}, which is hard to search.

For the above issues, there are some strategies to solve , however, there is no one which can solve   all the above problems.

The retrieve information by directory search engine  such as Yahoo  is more precise, but it takes too much human operations, and it is difficult to automatically handle.

Distributed search engines are able to break the  the ceiling of computer power and economic feasibility.   There are two kinds of distributed search engines. 1. the tasks for traditional search engine are scheduled to cluster computers in parallel computation such as Google or Baidu. 2.  The other uses P2P technologies.  However,  the un-structural P2P technology such as Freenet\cite{Clarke2000} using flooding way has shortage of heavy traffic and un-guaranteed search ;  the structural P2P technology using DHT such as Chord \cite{Stoica2001} and Pastry \cite{Rowstron2001} loses semantic meanings. These two kinds of P2P technologies are difficult to implement distributed search engines.

Most search engines do not retrieve information behind  of NAT computers.  There are lots of users who are willing to share there information in their computers which are behind of NAT.

The goals of DScloud platform\cite{DScloudplatform} are to solve the above problems. DScloud platform  based on   VIRGO \cite{LHuangVirgo}\cite{LHuangP2P}   P2P
network  which  hybrids structural and unstructured P2P technologies by merging n-tuple replicated virtual tree structured route nodes and
randomly cached un-structured route nodes( LRU and MinD). In VIRGO framework , the nodes are classified as multi-layer hierarchical catalogue domains according to their contents.  The search engine service of DScloud platform  includes automatical  article  digest, automatical   domain classification, local search engine and distributed search service.

The following structure of this paper is as follows: section 2 presents overview of Semantic P2P Networks;  section 3 describes
framework for Distributed Search Engine based on Semantic P2P Networks and its protocols;   section 4 presents a primary
design of Distributed Search Engine;  section 5 discuss security and unlawful search concerns and finally we give
conclusions.

\section{Overview  of Semantic P2P Networks   }

   Semantic P2P networks are based on   VIRGO \cite{LHuangVirgo}  \cite{agentH} which is a domain-related hierarchical structure
 hybridizing  un-structural P2P and structural P2P technology. VIRGO
 consists of prerequisite virtual group tree, and cached connections.
 Virtual group tree is similar to VDHA \cite{VDHA2003},
 but with multiple gateway nodes in every group. Virtual group tree
 is virtually hierarchical, with one root-layer, several middle-layers,
 and many leaf virtual groups.  Each group has  N-tuple gateway nodes.
 In VIRGO network, random connections cached  in a node's route table
 are maintained.  These cached connections  make VIRGO a distributed
 network, not just a virtual tree network like VDHA. With random cached
 connections, the net-like VIRGO avoids overload in root node in virtual
 tree topology, but keeps the advantage of effective message routing
 in tree-like network.

\begin{figure}[h]
 \begin{center}
 \setlength{\unitlength}{1cm}
 \begin{picture}(10,6)
 \includegraphics[scale=.35]{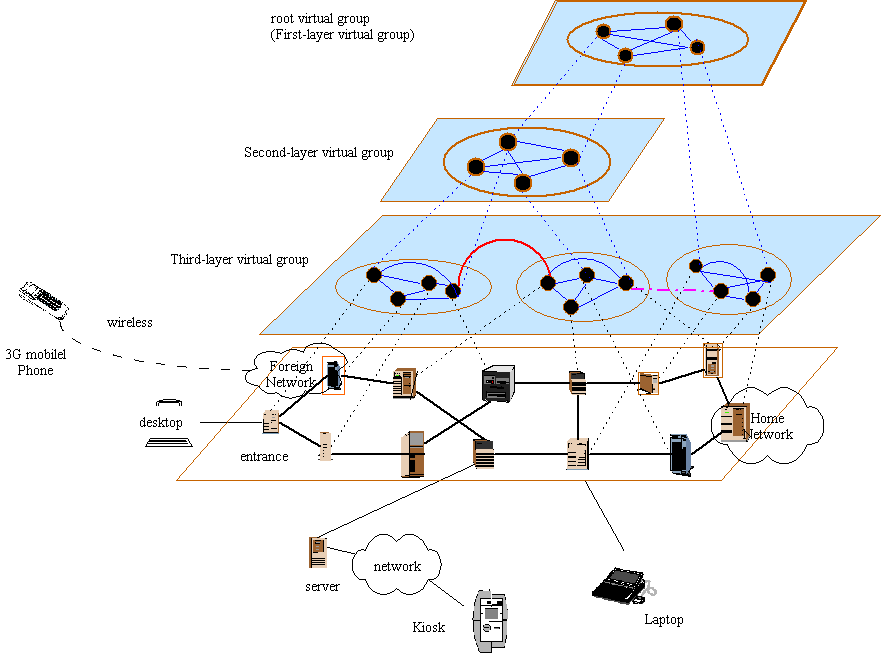}
 \end{picture}
 \caption{ Two\_tuple  Virtual Hierarchical Tree Topology} \label{1}
 \end{center}
\end{figure}

Fig. 1 shows two-tuple virtual hierarchical tree topology (the
nodes in different layers connected with dash line are actually
the same node). In Figure 1, from the real network, three virtual
overlay groups are organized. From these virtual groups two nodes
per virtual group are chosen to form the upper layer virtual
group.

In semantic P2P networks, the information are classified according to its domains, and stored in nodes which labeled as  related classification domains.

DScloud platform (http://www.yvsou.com) is based on semantic P2P network, in which the users can create domains according to standard classifications of industries, popular  classifications or according to their own classifications as fig.2 shows.  In DScloud platform  , the root domain is ALL, and  users can create new sub domains according to their authority. Users can join the sub-domains which they are interested in.

\begin{figure}[t]
 \begin{center}
 \setlength{\unitlength}{1cm}
 \begin{picture}(10,8)
 \includegraphics[scale=.30]{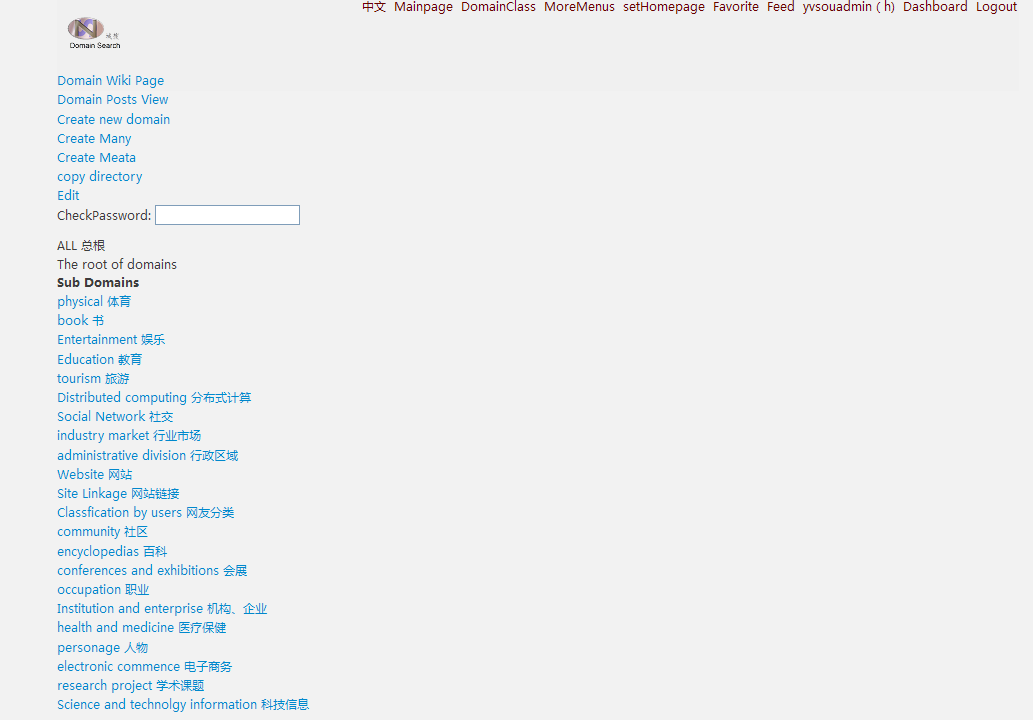}
 \end{picture}
 \caption{Domain classifications of DScloud platform } \label{1}
 \end{center}
\end{figure}

\section{Framework for Distributed Search Engine   based on
Semantic P2P Networks}

      A hierarchical virtual group tree is built according to domains. Users' computers   join the related groups. Each node has browser, search engine and Web Server besides semantic P2P network protocol implementation.  Framework for Distributed Search Engine   based on
Semantic P2P Networks is as fig.3 shows.

\begin{figure}[h]
 \begin{center}
 \setlength{\unitlength}{1cm}
 \begin{picture}(10,7)
 \includegraphics[scale=.30]{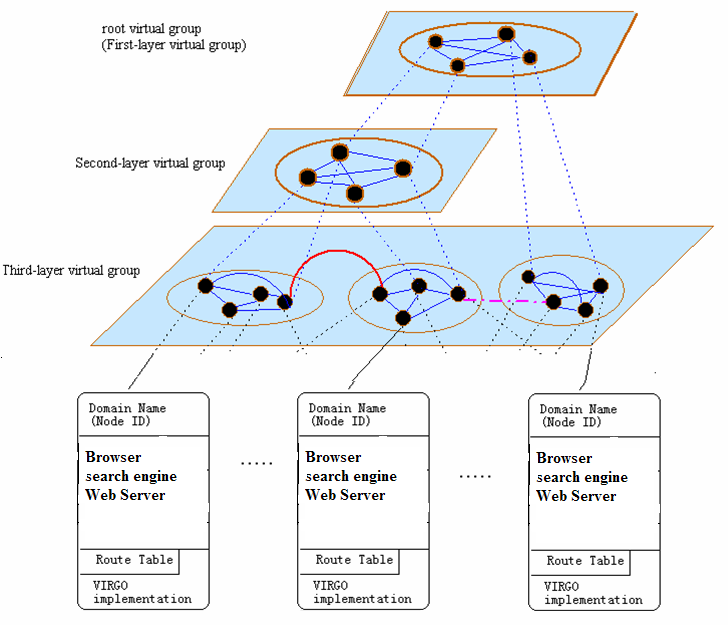}
 \end{picture}
 \caption{ Framework for Distributed Search Engine   based on
Semantic P2P Networks} \label{2}
 \end{center}
\end{figure}

In fig.3,  each node implements local search engine which indexes node's local information in node's computer.

\subsection{Protocols of Distributed Search Engine   based on
Semantic P2P Networks}

Unlike  key words as Google, here the search input format is keywords@domain. For example, if we want to search all information which is related with the key word- Windows 10 and within the domain-  all.education.undergraduated course.operating systems , so the input will be as :  "Windows 10@all.education.
undergraduated course.operating systems" \cite{patent1}\cite{patent2}\cite{patent3}.

\begin{verbatim}


Step 1  local VIRGO node forwards QUERY  MESSAGE to nearer
        host which joined  the same domain as the domain
        part of the input.

Step 2  the nearer host forwards QUERY MESSAGE to even
        nearer host until the destination host
        is found.

Step 3  the destination host retrieve information in the
        local machine using key word as the same of
        keyword part of the input;

Step 4  the destination host sends RESULT MESSAGE to
        original node;

Step 5  original node collects  all RESULT MESSAGEs .



\end{verbatim}

\section{Prototype Implementation of  Distributed Search Engine }

The local distributed search engine uses lucene \cite{lucene} to index  and search keywords.  The various kinds of documents in the directories under the root directory are extracted by various tools as shown  in  fig.4.

\begin{figure}[h]
 \begin{center}
 \setlength{\unitlength}{1cm}
 \begin{picture}(10,8)
 \includegraphics[scale=.30]{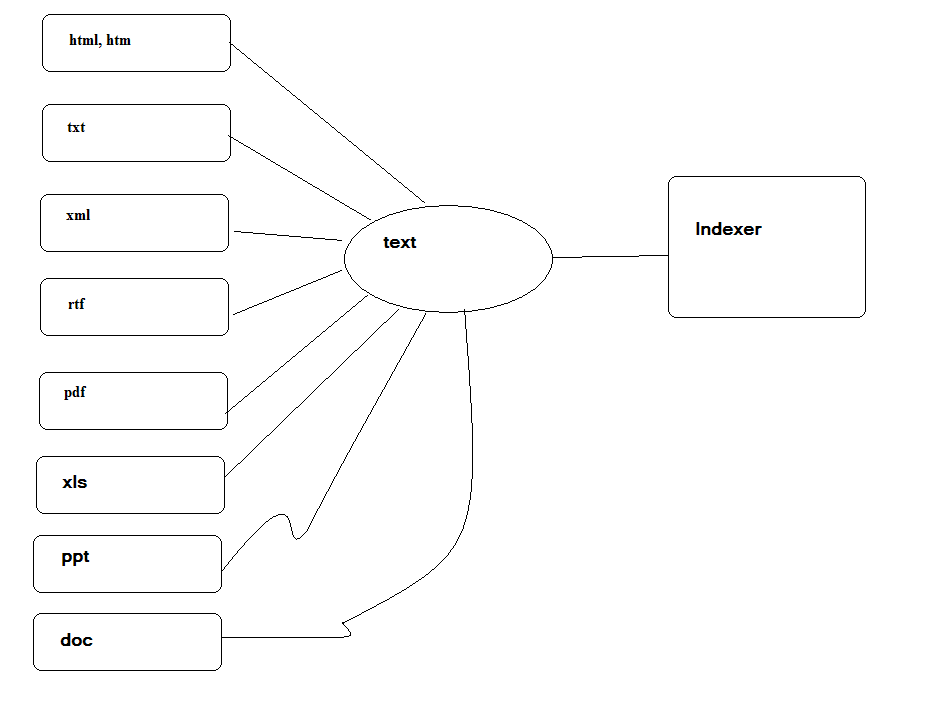}
 \end{picture}
 \caption{ Extract text from various kinds of documents} \label{2}
 \end{center}
\end{figure}

The steps of search process of the prototype is as the following:

\begin{verbatim}
Step 1  local browser  send QUERY MESSAGE to Local
        host web server

Step 2  local host  web server parses  QUERY MESSAGE and forwards
        the commands to  local VIRGO node.

Step 3  local VIRGO node forwards QUERY MESSAGE to nearer host
        with destination host which joined the same domain as
        domain part of the QUERY MESSAGE.

Step 4  the nearer host forwards  QUERY MESSAGE to even nearer
        host until the destination host is found.

Step 5  the destination host retrieve information using keyword
        in  QUERY MESSAGE in the local machine;

Step 6  the destination host sends RESULT MESSAGE to
        original node;

Step 7  original node collects  all RESULT MESSAGEs   ;

Step 8  original node sends all information to its local
        web server ;

Step 9  local  web server of original node sends all information
        to  its local browser;

Step 10  original node's local browser  displays results;

Step 11  When the original node's  local browser  clicks the links
         of the information, it uses UDP protocol to get the
         documents through the communications,and  displays the
         documents in the browser;


\end{verbatim}

 The prototype project of  distributed search engine   based on
Semantic P2P Networks which will be published in DScloud platform is shown in fig.5.

\begin{figure}[h]
 \begin{center}
 \setlength{\unitlength}{1cm}
 \begin{picture}(10,9)
 \includegraphics[scale=.30]{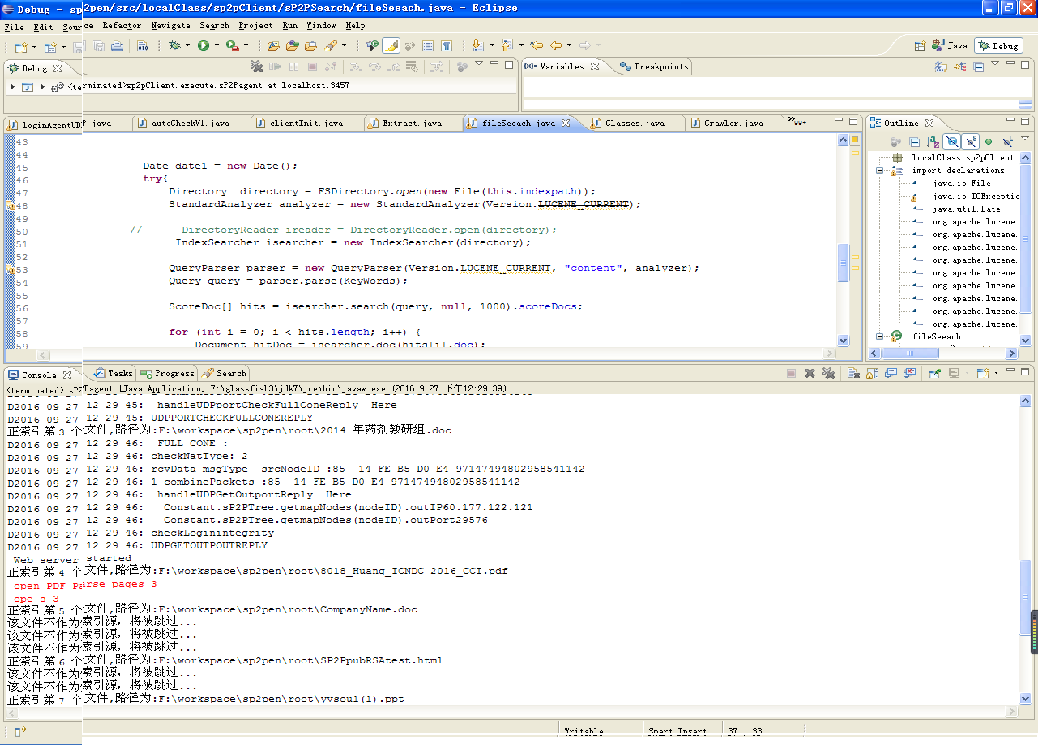}
 \end{picture}
 \caption{ Prototype project of  Distributed Search Engine   based on
Semantic P2P Networks} \label{2}
 \end{center}
\end{figure}

If the node wants to share files, firstly it joins a domain in the http://www.yvsou.com, then binds its user account with this machine, and puts the files being shared into the root directory .  The node needs to run sp2pn.jar.  If the other node wants to see the files of this node, firstly, it must run sp2pn.jar in its local machine, and lists all directories of the remote node as fig.6 shown.

\begin{figure}[h]
 \begin{center}
 \setlength{\unitlength}{1cm}
 \begin{picture}(10,5)
 \includegraphics[scale=.30]{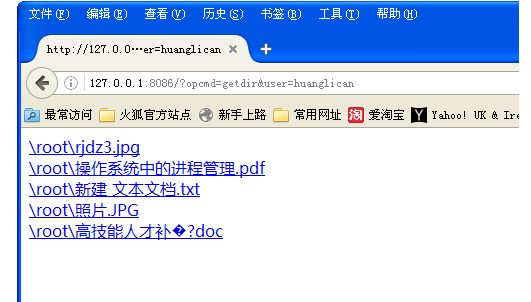}
 \end{picture}
 \caption{ Get directory from remote node} \label{2}
 \end{center}
\end{figure}

When we click the links of shared files in fig.5, then the file will be displayed in browser as shown in fig.6.
\begin{figure}[h]
 \begin{center}
 \setlength{\unitlength}{1cm}
 \begin{picture}(10,6)
 \includegraphics[scale=.30]{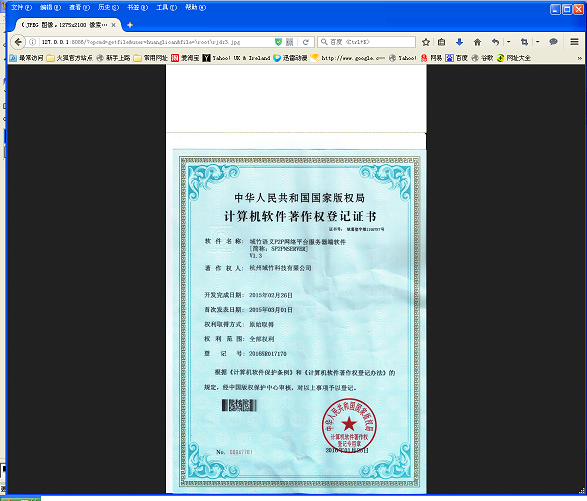}
 \end{picture}
 \caption{ Display file from remote node} \label{2}
 \end{center}
\end{figure}

\section{Privacy , Security and Unlawful Search Concerns }
 Although distributed search engine we presented may be slower  responsive than Google or Baidu, users may still use it for its full information retrieved. However, people may worry about the  privacy, security of  their computers. And government and companies worry about unlawful information transportation such as porn  information, piracy information.  In our prototype, we limit the working directory and its sub-directories as search directory. The search engine only retrieve these directories.  We also use trusted computing technologies to keep the integrity of the program to defend hacks' stealing and cheating information.  The government and companies can inform against the users who support  unlawful information  because  all users who support information have account. We can easily track and block the users with supporting unlawful information.

\section{Conclusion}
   We here present  framework and prototype of implementation  of a distributed search engine based on semantic P2P networks.  This framework has potential  to solve the problems such as  less precise   search results ,   the ceiling of computer power and economic feasibility and information retrieve of  local computers which are behind NAT.  Although distributed search engine we presented may be slower responsive than Google or Baidu, users may still use it for its full information retrieved. We also discuss the  strategy to solve the problems of the   privacy, security  and  unlawful information.

\section*{Acknowledgements}
The paper is supported by the project "e-business Plateform via Virtual Community constructed from Semantic P2P Networks" supported by Zhejiang future science and Technology City, and ¡°Hangzhou Qinglan
Plan--scientific and technological creation and development
(No.20131831K99¡± supported by  Hangzhou scientific and technological
committee. The software copyrights is owned by Hangzhou
Domain Zones Technology Co., Ltd. and Domain Search Networking Technology Co., Ltd., and Chinese patent
applied is owned by Hangzhou Domain Zones Technology
Company and Domain Search Networking Technology Co., Ltd..

\end{document}